\documentclass[aip, preprint]{revtex4-1}
\usepackage{graphicx}
\usepackage{textcomp} 
\setcitestyle{authoryear,round,aysep={,},yysep={,}}
\usepackage[none]{hyphenat}

\begin{document}

\title[Measurement Error in Atomic-Scale STEM-EDS]{Measurement Error in Atomic-Scale STEM-EDS Mapping of a Model Oxide Interface}

\author{Steven R. Spurgeon}
\affiliation{Physical and Computational Sciences Directorate, Pacific Northwest National Laboratory, Richland, Washington 99352, USA}

\author{Yingge Du}
\affiliation{Physical and Computational Sciences Directorate, Pacific Northwest National Laboratory, Richland, Washington 99352, USA}

\author{Scott A. Chambers}
\email{sa.chambers@pnnl.gov}
\thanks{Phone:+1-509-371-6517; Fax:+1-509-371-6066}
\affiliation{Physical and Computational Sciences Directorate, Pacific Northwest National Laboratory, Richland, Washington 99352, USA}

\date{\today}

\begin{abstract}

With the development of affordable aberration-correctors, analytical scanning transmission electron microscopy (STEM) studies of complex interfaces can now be conducted at high spatial resolution at laboratories worldwide. Energy-dispersive X-ray spectroscopy (STEM-EDS) in particular has grown in popularity, since it enables elemental mapping over a wide range of ionization energies. However, the interpretation of atomically-resolved data is greatly complicated by beam-sample interactions that are often overlooked by novice users. Here we describe the practical factors---namely, sample thickness and the choice of ionization edge---that affect the quantification of a model perovskite oxide interface. Our measurements of the same sample in regions of different thickness indicate that interface profiles can vary by as much as 2--5 unit cells, depending on the spectral feature. This finding is supported by multislice simulations, which reveal that on-axis maps of even perfectly abrupt interfaces exhibit significant delocalization. Quantification of thicker samples is further complicated by channeling to heavier sites across the interface, as well as an increased signal background. We show that extreme care must be taken to prepare samples to minimize channeling effects and argue that it may not be possible to extract atomically-resolved information from many chemical maps.

\end{abstract}

\maketitle

\section*{Introduction}

Interfaces control the behavior of a variety of emergent properties in oxides, ranging from electron gas formation \citep{Chakhalian2014} to ferroelectricity \citep{Mannhart2010}. While great strides have been made in the precision synthesis of atomically-sharp thin film heterostructures \citep{Martin2010}, interface charge \citep{Nakagawa2006}, strain \citep{SankaraRamaKrishnan2014}, and entropy can all drive film-substrate intermixing that can greatly affect properties. With the development of commercial spherical aberration (C$_\textrm{s}$) corrected microscopes, analytical scanning transmission electron microscopy (STEM) is now increasingly used to characterize nanoscale interfaces \citep{Krivanek2008}. C$_\textrm{s}$-correction has enabled the large convergence angles, small probe sizes, and high probe currents needed for efficient chemical mapping of individual atomic columns \citep{Lu2013, Allen2012, DAlfonso2010, Muller2008}. Complementary analysis using energy-dispersive X-ray spectroscopy (STEM-EDS) and electron energy loss spectroscopy (STEM-EELS) now permits measurements of composition and chemistry at the \AA ngstr\"{o}m scale. Moreover, cost reductions have now made aberration-correctors more attainable than ever, leading to their widespread adoption by universities and laboratories. While the site-specific nature of these techniques can offer rich insight into local interface environments, the interpretation of the resulting data is far from simple and still poorly understood, as discussed in a recent case study of the perovskite SrTiO$_3$ (STO) \citep{Kothleitner2014}. A complex array of physical processes, including beam broadening and channeling effects \citep{Oxley2007}, can lead to serious misinterpretations of chemical maps. Channeling is particularly problematic, since it tends to occur when imaging along low-order zone axes commonly used for atomic-scale imaging; in this case, the strong Coulombic interaction between the electron probe and the atoms in the crystal focuses the probe intensity along columns, complicating the analysis of ionization signals \citep{Lugg2014}.

The newly developed technique of atomic-column STEM-EDS mapping exemplifies the challenges associated with the quantification of high-resolution analysis of crystalline materials. In one of the first demonstrations of atomic-scale STEM-EDS mapping of STO \citep{DAlfonso2010}, D'Alfonso \textit{et al.} argued that the localization of X-ray scattering potentials can lead to a directly interpretable chemical map analogous to high-angle annular dark field (STEM-HAADF) imaging \citep{Allen2012}. However, subsequent work has shown that the contribution of thermally scattered electrons can affect contrast in both EDS and EELS maps \citep{Forbes2012}, while studies of interfaces have revealed that the apparent atomic column size is affected by probe channeling, which directly depends on thickness \citep{Lu2014, Lu2013}. Similar difficulties are encountered in STEM-EELS mapping, where image contrast reversals have been observed that are attributed to off-column channeling of the probe \citep{Wang2008, Oxley2007}. Probe broadening due to a finite sample thickness can also complicate the analysis of diffuse interfaces. The few quantitative studies conducted to date have shown that the characterization of unknown structures is difficult \citep{Kotula2012}, generally requiring extensive modeling \citep{Neish2015}, and \textit{a priori} sample information \citep{Kothleitner2014}. Promising recent work by Chen \textit{et al.} has shown that it may be possible to quantify EDS maps on an absolute scale by combining thin samples with rigorous simulations \citep{Chen2016, Chen2015}.

A study of several key parameters in STEM-EDS measurements is needed to inform the growing community and raise awareness of potential sources of error. Here we consider how sample thickness and the choice of X-ray ionization edge influence measurements of interface composition and mixing in a model perovskite oxide interface. We find that measurements of the interface width can vary widely from 2--5 unit cells and that spectral components can exhibit different thickness dependencies, leading to variations in measurement error for each species. Moreover, channeling of the delocalized signal to heavier sites across the interface can complicate direct quantification of peak areas. Our study highlights important practical considerations for atomic-scale STEM-EDS mapping of interfaces, emphasizing the need for both extremely thin samples and supporting simulations to accurately interpret experimental data. We caution against direct interpretation of chemical maps absent these qualifications.

\section*{Materials and Methods}

We have selected a model system consisting of a 30 nm-thick La$_{0.88}$Sr$_{0.12}$CrO$_3$ ($p$-type LSCO) thin film deposited onto a 0.1 Wt\% Nb:SrTiO$_3$ (001) ($n$-type Nb:STO) substrate using molecular beam epitaxy \citep{Zhang2015}, as illustrated in Figure \ref{eds_map}(a). This system is intriguing for its potential use as an all-perovskite transparent $p-n$ junction, as will be reported elsewhere. Predictive control of device performance depends on a clear understanding of the interface structure to preserve the band offset between the two layers. TEM samples were prepared using a standard lift out method on an FEI Helios DualBeam focused ion beam (FIB) microscope. To generate a range of sample thicknesses, a wedge-shaped sample was prepared at 4--7$^{\circ}$ incidence angle, using ion beam energies of 2--30 keV. STEM measurements were performed at 200 keV using an aberration-corrected JEOL ARM-200CF microscope equipped with a JEOL Centurio silicon drift detector (quoted solid angle of 0.98 sr) for EDS analysis. All images were acquired along the [100] zone-axis with a $\sim 1$ \AA \, probe size and a 27.5 mrad convergence semi-angle, yielding an approximate probe current of $\sim 130$ pA. STEM-HAADF images were acquired with 90--370 mrad inner-outer collection angles, respectively, while STEM-EDS maps were acquired in multiple regions using the Thermo Noran System 7 software, with an approximate instantaneous pixel dwell time of $50$ \textmu s px$^{-1}$, an effective total dwell time of 9--27 ms px$^{-1}$, and a total collection time of 5--17 min. Maps were processed for net counts, with a background removal and multiple linear least squares fit of reference spectra to deconvolve overlapping peaks. STEM-EELS zero-loss peak thickness measurements indicate that the measured regions are approximately 28, 33, 50, 66, 70, and 75 nm-thick.

Multislice simulations were conducted using the quantum excitation of phonons (QEP) model \citep{Forbes2010}, which accounts for beam channeling effects and allows for separation of elastic and thermal electrons, in contrast to the frozen phonon model. Our simulations used a supercell consisting of six unit cells of TiO$_2$-terminated SrTiO$_3$ (STO) interfaced with six unit cells of La$_{0.88}$Sr$_{0.12}$CrO$_3$ (LSCO). STEM-EDS ionization maps were simulated using the \textmu STEM v4.5 software package \citep{Allen2015} with a $10 \times 1$ supercell tiling and $2500 \times 2160$ px grid sampling for 10, 28, 50, and 100 nm-thick crystals. The total simulation time was 25 days. Using these parameters we are able account for electrons scattered to a maximum angle of approximately 300 mrad. X-ray absorption has not been included in these models, which may impact the results for thicker samples. Crystal models were converted into the appropriate input format using the XTL-Converter program \citep{SpurgeonNov2015}. Chen \textit{et al.} have shown that the size of the necessary Gaussian finite source correction can be broadened by multiple-frame averaging \citep{Chen2016}; therefore, to determine the most appropriate source size we calculated a range of corrections and compared them to the experimental data. This procedure yielded a FWHM = 0.19 nm as the best fit.

\section*{Results and Discussion}

\begin{figure}
\includegraphics[width=\textwidth]{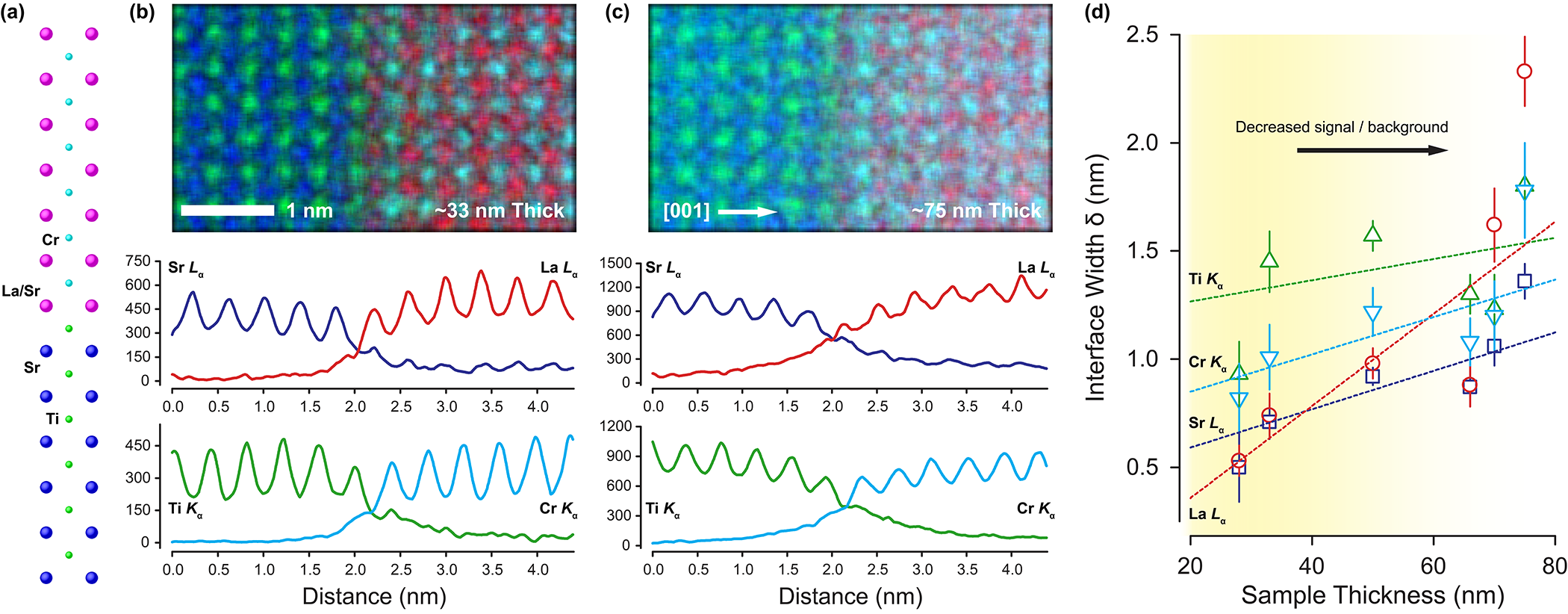}
\caption{(a) Illustration of the STO / LSCO crystal structure. (b--c) Unfiltered composite STEM-EDS maps and corresponding $A$- and $B$-site net X-ray count line profiles for $\sim$33 and $\sim$75 nm-thick STO / LSCO interfaces, respectively. The line profiles have been averaged in the plane of the maps. (d) Interface width as a function of sample thickness for each edge. \label{eds_map}}
\end{figure}

Figures \ref{eds_map}(b--c) show composite STEM-EDS maps and corresponding averaged line profiles for Sr $L_\alpha$ (1.806 kV), La $L_\alpha$ (4.650 kV), Ti $K_\alpha$ (4.508 kV), and Cr $K_\alpha$ (5.412 kV) taken from $\sim$33 and $\sim$75 nm-thick regions, respectively. The profiles were extracted from the net X-ray counts maps, averaged in the film plane, and smoothed using an adjacent averaging filter. For comparison to simulation, we propose the use of a logistic fit to the extrema on either side of the interface; we define the interface width ($\delta$) as the difference of 90\% and 10\% of the signal maxima. Using this procedure we measure the interface width as a function of sample thickness, as shown in Figure \ref{eds_map}(d). Even though all measurements are conducted on the same crystal, we find that there is a clear increase in the apparent interface width in thicker mapping regions. $\delta_\textrm{La}$ by far exhibits the largest change in interface width, increasing nearly fourfold by $\sim 1.8$ nm moving from 28 to 75 nm-thick regions. $\delta_\textrm{Sr}$ and $\delta_\textrm{Cr}$ increase by $\sim 0.86$ and $\sim 0.96$ nm, respectively, while $\delta_\textrm{Ti}$ increases by $\sim 0.87$ nm. Depending on the species, these values correspond to nearly 2--5 unit cells of measurement error. We find that the absolute value of $\delta_\textrm{Ti}$ is generally larger than the other species in the thinnest regions; significant Ti interdiffusion is not unexpected in this system and may actually promote a more stable interface \citep{Colby2013}. We note that atomic force microscope (AFM) maps of the substrate surface show that at most one step can be contained in the foil along the beam direction, which could also affect the absolute value of the interface width in any MBE-grown samples. In addition to the broadening of interface profiles, we also find that the peak-to-background ratios of the profiles significantly decrease with increasing thickness, as evidenced by comparing Figures \ref{eds_map}(b--c). These results are troubling and suggest that the accuracy of interface measurements, as well as the quantification of resulting peak areas, can greatly depend on sample thickness.

\begin{figure}
\includegraphics[width=\textwidth]{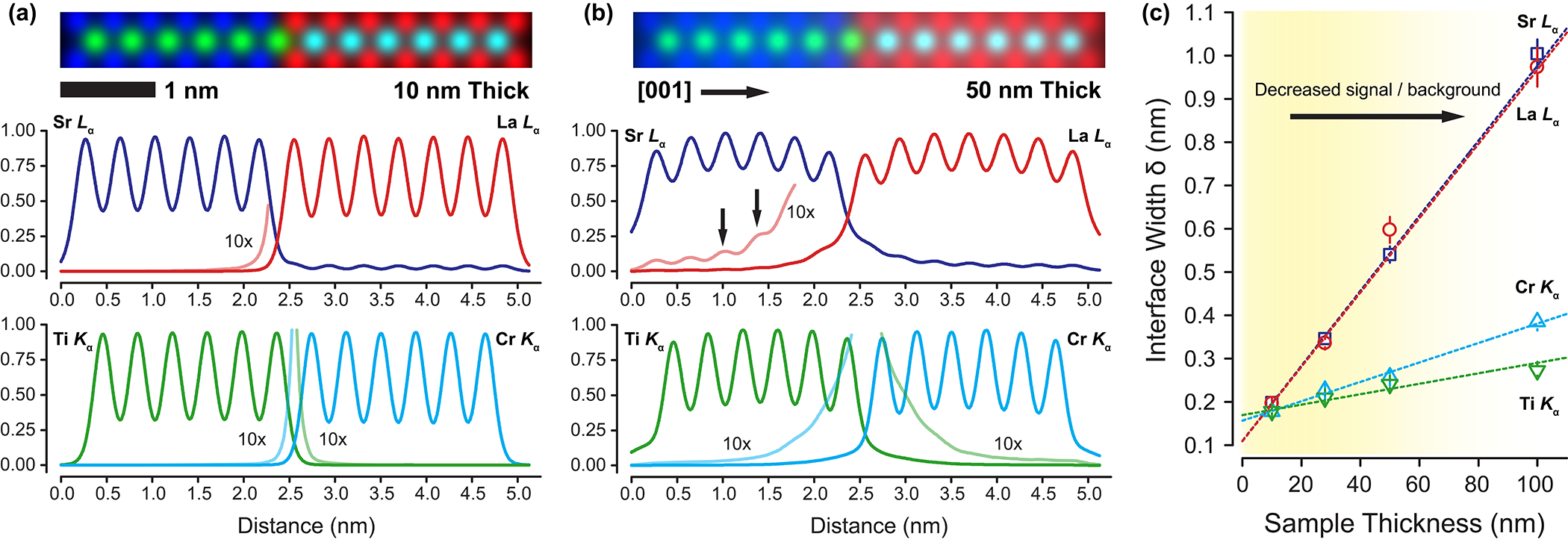}
\caption{(a--b) Simulated STEM-EDS ionization maps and corresponding $A$- and $B$-site line profiles for abrupt 10 and 50 nm-thick STO / LSCO interfaces, respectively. The line profiles have been normalized to the signal maxima and are inset with a $10 \times$ magnified signal to emphasize channeling effects. (c) Interface width as a function of model thickness for each edge.\label{multislice}}
\end{figure}

To gain insight into the effects of delocalization and channeling on the resulting ionization maps, we have performed multislice simulations, systematically varying the thickness of a model interface. We consider an ideal abrupt STO / La$_{0.88}$Sr$_{0.12}$CrO$_3$ interface, thereby avoiding the complications of modeling different interface geometries. We have simulated ionization maps for 10, 28, 50, and 100 nm-thick samples; selected ionization maps and line profiles for the 10 and 50 nm-thick samples are shown in Figures \ref{multislice}(a-b), respectively, alongside a plot of interface width \textit{versus} crystal thickness in Figure \ref{multislice}(c). A comparison of the maps in Figures \ref{multislice}(a-b) shows that the 50 nm-thick sample exhibits a higher background than the 10 nm-thick sample, as indicated by the diffuse bands of contrast between lattice sites. Figure \ref{multislice}(a) shows line profiles taken from both the $A$- and $B$-site columns of the 10 nm-thick sample; here we find that the ionization signal of all four edges is strongly localized to their respective atomic columns, with a low background between peaks. This case corresponds to the ultrathin limit previously described by Lu \textit{et al.}  \citep{Lu2013}, where the electron wave is, in principle, directly convolved with the local EDS ionization potential and is less affected by beam channeling. Using the same logistic fitting procedure we measure the following interface widths: $\delta_\textrm{Sr} = 0.20$, $\delta_\textrm{Ti} = 0.18$, $\delta_\textrm{La} = 0.20$, and $\delta_\textrm{Cr} = 0.18$ nm (all $\pm \, 0.01$ nm). While our simulations cannot account for all factors, such as thickness or strain fluctuations across the interface, we estimate that this result represents the lower limit of intermixing measurements in an extremely thin and abrupt interface; even in the case of such an ideal sample, there is still some artificial intermixing.

Line profiles from the 50 nm-thick model, shown in Figure \ref{multislice}(b), reveal very different behavior from the 10 nm-thick case. The background between lattice sites has significantly increased, resulting from the redistribution of probe intensity within the sample caused by channeling. Owing to the intrinsic delocalization of lower energy edges, the background for the Sr $L_\alpha$ and La $L_\alpha$ edges is nearly 75\% of the signal maximum, while the background for the Ti $K_\alpha$ and Cr $K_\alpha$ edges is less than 50\% of the maximum. The choice of ionization edge may therefore affect the spatial distribution of EDS ionization potential, as has been observed in STEM-EELS mapping \citep{Wang2008}, as well as the accuracy of composition quantification based on peak area fitting \citep{Lu2014}. Furthermore, the $A$-site line profiles show that a substantial fraction ($10-15\%$) of the signal can be delocalized across even a perfectly abrupt interface. We find that the La $L_\alpha$ signal is delocalized across the interface to Sr positions, as marked by the arrows in Figure \ref{multislice}(b); a similar, albeit much reduced, effect is seen for the $B$-sites. This delocalization and subsequent channeling to heavier sites can arise from both thermally and elastically scattered electrons, which are then able to go on to ionize other atoms \citep{Forbes2012}. As expected, delocalization results in a sizable increase in the apparent interface width, yielding: $\delta_\textrm{Sr} = 0.54 \pm 0.02$, $\delta_\textrm{Ti} = 0.26 \pm 0.01$, $\delta_\textrm{La} = 0.60 \pm 0.03$, and $\delta_\textrm{Cr} = 0.24 \pm 0.01$ nm. We again find that delocalization is much more pronounced for the Sr $L_\alpha$ and La $L_\alpha$ signals, tripling $\delta$ from $0.20$ to $0.54-0.60$ nm. This width corresponds to more than a unit cell of artificial intermixing (measurement error), even in the case of a perfectly abrupt interface structure. On the other hand, the Ti $K_\alpha$ and Cr $K_\alpha$ signals show only a negligible increase in artificial interface width compared to the 10 nm-thick case.

A comparison to our experiments shows that, even though the measured profiles are wider than the simulations, there is a consistent trend toward an artificially broadened interface in thicker regions. In agreement with our simulations, we find that the La $L_\alpha$ and Sr $L_\alpha$ signals exhibit the largest increase in broadening, but the other species exhibit similar trends. Most importantly, we observe that the absolute value of the interface width can vary as much as 2--5 unit cells, making it difficult to extract a meaningful picture of the interface in thicker mapping regions. While such regions may only be 40--50 nm thick, that is more than sufficient to introduce noticeable measurement errors. In summary, our results highlight two main difficulties faced in the quantification of interfaces: electron probe channeling can greatly alter the on- and off-column signal, leading to sizable artificial intermixing over several unit cells; further, the measurement error also depends on the intrinsic delocalization associated with a chosen ionization edge.

\section*{Conclusions}

In light of these results, great care must be taken to conduct meaningful studies of interfaces using the aberration-corrected STEM-EDS technique. We find that delocalization effects are minimized in the thinnest regions and that the best results are obtained below 25--30 nm; however, this imposes a severe limit on sample preparation and signal collection times for accurate quantification. We also emphasize that multislice simulations should be conducted for each system and ionization edge of interest to aid the interpretation of experimental data. Future GPU-accelerated computing will allow us to produce more rigorous models to better simulate real-world interfaces and help us account for complex delocalization processes. For now it is imperative that novice users understand the limitations of chemical mapping and that steps are taken to prepare sufficiently thin samples, depending on the level of accuracy desired.

\section*{Acknowledgments}

S.R.S. thanks Drs. Scott Findlay, Yuanyuan Zhu, Despoina Kepaptsoglou, and Matthew Olszta for insightful discussions. This work was supported by the U.S. Department of Energy (DOE), Office of Science, Division of Materials Sciences and Engineering under award \#10122. This work was performed in the Environmental Molecular Sciences Laboratory (EMSL), a national science user facility sponsored by the DOE Office of Biological and Environmental Research (BER) and located at Pacific Northwest National Laboratory (PNNL), operated for DOE by Battelle. Multislice calculations were performed using PNNL Institutional Computing.

\newpage

\bibliographystyle{MandM}
\bibliography{References}

\end{document}